\RequirePackage{silence}
\documentclass[fleqn,usenatbib,useAMS]{mnras} 
\usepackage{graphicx}
\usepackage{amsmath}

\usepackage{amsfonts}
\usepackage{float}
\usepackage{bm}
\usepackage[perpage,symbol*]{footmisc}
\usepackage{ae,aecompl}
\usepackage{array}
\usepackage{soul}
\usepackage{mathtools}
\usepackage{multirow}

\newcommand{\appropto}{\mathrel{\vcenter{
			\offinterlineskip\halign{\hfil$##$\cr 
				\propto\cr\noalign{\kern2pt}\sim\cr\noalign{\kern-2pt}}}}}

\makeatletter

\newcommand{\Rmnum}[1]{\expandafter\@slowromancap\romannumeral #1@}
\makeatother

\title{The External Field Dominated Solution In QUMOND \& AQUAL: Application To Tidal Streams}
\author[Indranil Banik \& Hongsheng Zhao]{Indranil Banik$^{1}$\thanks{Email: ib45@st-andrews.ac.uk (Indranil Banik), \newline $~~~~~~~~~~~~~~~$hz4@st-andrews.ac.uk (Hongsheng Zhao)}, Hongsheng Zhao$^{1}$\\
$^{1}$Scottish Universities Physics Alliance, University of St Andrews, North Haugh, St Andrews, Fife, KY16 9SS, UK}

\pubyear{2018}
\begin{document}
\label{firstpage}
\pagerange{\pageref{firstpage}--\pageref{lastpage}}

\maketitle

\begin{abstract}

The standard $\Lambda$CDM paradigm seems to describe cosmology and large scale structure formation very well. However, a number of puzzling observations remain on galactic scales. An example is the anisotropic distribution of satellite galaxies in the Local Group. This has led to suggestions that a modified gravity theory might provide a better explanation than Newtonian gravity supplemented by dark matter. One of the leading modified gravity theories is Modified Newtonian Dynamics (MOND). For an isolated point mass, it boosts gravity by an acceleration-dependent factor of $\nu$.

Recently, a much more computer-friendly quasi-linear formulation of MOND (QUMOND) has become available. We investigate analytically the solution for a point mass embedded in a constant external field of $\bm{g}_{ext}$. We find that the potential is $\Phi = - ~ \frac{GM \nu_{ext}}{r}\left(1 + \frac{K_0}{2} \sin^2 \theta \right)$, where $r$ is distance from the mass $M$ which is in an external field that `saturates' the $\nu$ function at the value $\nu_{ext}$, leading to a fixed value of $K_0 \equiv \frac{\partial Ln ~ \nu}{\partial Ln ~ g_{ext}}$. In a very weak gravitational field $\left(\left| \bm{g}_{ext} \right| \ll a_0 \right)$, $K_0 = -\frac{1}{2}$. The angle $\theta$ is that between the external field direction and the direction towards the mass.

Our results are quite close to the more traditional aquadratic Lagrangian (AQUAL) formulation of MOND. We apply both theories to a simple model of the Sagittarius tidal stream. We find that they give very similar results, with the tidal stream seeming to spread slightly further in AQUAL.

\end{abstract}


\section{Introduction}
\label{Introduction}

The standard $\Lambda$CDM paradigm \citep{LCDM_Proposal} still faces many challenges in reproducing galaxy scale observations \citep[for a recent review, see][]{MOND_Review}. Particularly problematic is the anisotropic distribution of satellite galaxies around Local Group galaxies, a question recently revisited in detail \citep{Pawlowski_2014}. A different analysis focusing on Andromeda came to similar conclusions \citep{Ibata_2014}.

The relevant observations for the Milky Way \citep{VPOS_Proper_Motions} and Andromeda \citep{Andromeda_Satellite_Plane} are difficult to repeat outside the Local Group because of the need to obtain 3D positions and velocities. However, it has recently been claimed that such structures are common at low redshift \citep{Ibata_2014_Nature}. This study later faced some criticism \citep{Cautun_2015}, but these concerns appear to have been addressed \citep{Ibata_2015}.

For the case of the Milky Way (MW) and Andromeda (M31), it appears very unlikely that these structures formed quiescently \citep{Pawlowski_2014, Ibata_2014}. Filamentary infall is considered unlikely because it would imply the satellites had very eccentric orbits, contrary to observations of the MW satellites \citep{Angus_2011}. Moreover, they would need to have been accreted long ago in order to give time to circularise their orbits via dynamical friction with the dark matter (DM) halo of the MW.

However, interactions between satellites and numerous DM halos that are thought to surround the MW \citep{Klypin_1999} would lead to the spreading out of any initially thin disk of satellites \citep{Klimentowski_2010}. Even if the number of subhaloes was smaller than predicted by $\Lambda$CDM, the triaxial nature of the potential would still cause dispersal over a timescale of ${\approx 5}$ Gyr \citep{Bowden_2013}. This is not true if the structure was aligned with an axis of symmetry of the potential, but such a perfect alignment seems unlikely in the only two galaxies observed at this level of detail.

One possibility is that an ancient interaction created the satellites as tidal dwarf galaxies \citep[TDGs, ][]{Kroupa_2005}. After all, we do see galaxies forming from material pulled out of interacting progenitor galaxies \citep[e.g. in the Antennae][]{Mirabel_1992}. This naturally leads to anisotropy because the tidal debris tend to be confined to the common orbital plane of the interacting progenitor galaxies.

Unlike the baryons, the DM must be pressure-supported, making it difficult to draw into dense tidal tails. As a result, TDGs should be free of DM \citep{Barnes_1992}. Thus, a surprising aspect of LG satellite galaxies is their high mass-to-light (M/L) ratios \citep[e.g.][]{McGaugh_2013}. These are calculated assuming dynamical equilibrium. Tides from the host galaxy can invalidate this assumption. However, tides in $\Lambda$CDM are likely not strong enough to do this \citep{McGaugh_2010}. Similar conclusions were drawn about TDGs near the Seashell galaxy \citep{Gentile_2007, Bournaud_2007}. With dark matter unlikely to be present in these systems, the high inferred M/L ratios would need to be explained by modified gravity.

The most widely investigated such theory is Modified Newtonian Dynamics \citep[MOND,][]{Milgrom_1983}. In this theory, the MW and M31 would have undergone an ancient close flyby ${\approx 9}$ Gyr ago \citep{Zhao_2013}. The thick disk of the MW would then be naturally explained as having formed due to this interaction. Indeed, recent work suggests a tidal origin for the thick disk \citep{Banik_2014}. Moreover, its age is consistent with this scenario \citep{Quillen_2001}.


In MOND, adding a constant external gravitational field $\bm{g}_{ext}$ to a system affects it non-trivially, unlike in Newtonian gravity. This is because MOND is a non-linear theory. In a rich galaxy cluster, the effects can be substantial \citep{Wu_2007}. The external field on the Local Group affects the motion of the MW and M31 because it is comparable to the relative MW$-$M31 acceleration at apocentre \citep{Zhao_2013}. In MOND, external fields determine the escape speed from systems such as the MW \citep{Famaey_2007}. Internal dynamics of satellite galaxies can also be affected by gravity from the host \citep[e.g.][]{Angus_2014}.

Our focus is on systems where the external field is dominant. In Section \ref{Sgr_tidal_stream}, we consider the tidal stream left behind by the disrupting Sagittarius (Sgr) dwarf spheroidal galaxy \citep{Lynden_Bell_1995}. This is modelled using the modified Lagrange Cloud Stripping procedure \citep{Gibbons_2014}. Gravity from Sgr is important, but the total gravitational field strength is generally dominated by the MW. This is especially true at or beyond the tidal radius: merely the difference in gravity from the MW between the centre of Sgr and its tidal boundary is comparable to the internal gravity of Sgr on this boundary.

We are mostly interested in the dynamics of tidal stream particles. These must lie beyond the tidal radius of Sgr. As the density likely falls off sharply beyond the tidal radius, we use a point mass model for Sgr. Thus, we set about solving the governing equations of MOND for a point mass in a dominating external field $\bm{g}_{ext}$. Our solution is invalid for distances from Sgr of
\begin{eqnarray}
	r ~\la ~ \frac{\sqrt{GM_{Sgr} a_0}}{g_{ext}} ~~\text{ where $g_{ext}$ is due to the MW}
\end{eqnarray}

This is because gravity from Sgr dominates sufficiently close to it, if it is treated as a point mass. However, if it is extended, then gravity due to Sgr would eventually start to \emph{decrease} as one got closer to its centre. In this case, it is possible for the external field to dominate everywhere. In Section \ref{Sgr_tidal_stream}, we assume that it does.

In what follows, when we refer to `the mass', we mean Sgr and not the MW. We use $M$ for the mass of a point-like object immersed in a constant external gravitational field $\bm{g}_{ext}$. In Newtonian gravity, the external field would have been $\bm{g}_{_{N, ext}}$. To reduce the likelihood of $~-$ sign errors, we prefer to work with $\bm{n} \equiv \nabla \Phi \equiv -\bm{g}$, where $\Phi$ is the potential.

\section{External Field Dominance In AQUAL}

Firstly, we review the derivation in the original aquadratic Lagrangian (AQUAL) formulation of MOND \citep{Bekenstein_Milgrom_1984}. This follows the work of \citet{Milgrom_1986}. We separate the gravitational field into the part due to the mass and the external field, making the governing equation
\begin{eqnarray}
	\label{AQUAL_Governing_Equation}
	\nabla \cdot \left[ \mu \left( \left| \bm{n} + \bm{n}_{ext} \right| \right) ~\left( \bm{n} + \bm{n}_{ext} \right) \right] ~&=&~ 4 \pi G \rho ~\text{ where  } \\
	\bm{n} + \bm{n}_{ext} ~&\equiv &~ \nabla \Phi
\end{eqnarray}

The boundary condition is that $n \to 0 $ at long range. Because of the external field, $\Phi \to n_{ext} z$ if we use a Cartesian system with its $z$-axis along $\bm{n}_{ext}$.

The function $\mu$ is acceleration-dependent and key to AQUAL. For gravitational field strengths $n \gg a_0$, we must recover Newtonian gravity, forcing $\mu \to 1$. For $n \ll a_0$, observations of galaxies require $\mu \to \frac{n}{a_0}$. We use the form
\begin{eqnarray}
	\mu = \frac{n}{n + a_0}
	\label{Equation_simple_mu}
\end{eqnarray}

This is called the simple $\mu$ function \citep{Famaey_Binney_2005}. It seems to work well with observations, especially of our own Galaxy \citep{Iocco_Bertone_2015}.

We now linearise Equation \ref{AQUAL_Governing_Equation}, noting that $n_{ext} \gg n$ in the region of interest. Thus, $\mu \approx \mu \left( n_{ext} \right) \equiv \mu_{ext}$.
\begin{eqnarray}
	&& \nabla \cdot \left[ \mu \left( \left| \bm{n} + \bm{n}_{ext} \right| \right) ~ \left( \bm{n} + \bm{n}_{ext} \right) \right]\\
	&=& \mu \nabla \cdot \bm{n} ~+~ \left( \left( \bm{n} + \bm{n}_{ext} \right) \cdot \nabla \right) \mu \\
	& \approx & \mu_{ext} \nabla \cdot \bm{n} ~+~ \left( \bm{n}_{ext} \cdot \nabla \right) \left[ \mu_{ext} + \mu'\bm{n}_z  \right] ~\text{ where } \\
	\mu' &\equiv & \left. \frac{\partial \mu}{\partial n} \right|_{n = n_{ext}}
\label{Equation_6}
\end{eqnarray}


At first order, $\left| \bm{n} + \bm{n}_{ext} \right|$ is only affected by the component of $\bm{n}$ parallel to $\bm{n}_{ext}$. As we only seek the first order Taylor expansion of $\mu$, we see that only $\bm{n}_z$ can much affect it. After taking out a common factor of $\mu_{ext}$, we get that
\begin{eqnarray}
	\label{AQUAL_linearised}
	\mu_{ext} \left( \nabla \cdot \bm{n} + L_0  \frac{\partial \bm{n}_z}{\partial z}\right) &\approx & 4 \pi G \rho ~~~\text{where}  \\
	L_0 &\equiv & \left. \frac{\partial Ln ~ \mu}{\partial Ln ~ n}  \right|_{n ~=~ n_{ext}}
\end{eqnarray}

In Cartesian co-ordinates, Equation \ref{AQUAL_linearised} reads
\begin{eqnarray}
	\frac{\partial^2 \Phi}{\partial x^2} + \frac{\partial^2 \Phi}{\partial y^2} + \left( 1 + L_0 \right) \frac{\partial^2 \Phi}{\partial z^2} ~=~ \frac{4 \pi G \rho}{\mu_{ext}}
\end{eqnarray}

This can be reduced to a rescaled version of the normal Poisson equation if we set $z \to z' \equiv \frac{z}{\sqrt{1 + L_0}}$ but leave $x$ and $y$ unaltered. Thus, in the rescaled co-ordinates, we get that
\begin{eqnarray}
	\nabla'^2 \Phi ~=~ \frac{4 \pi G \rho}{\mu_{ext}}
\end{eqnarray}

As the equations are now linear, we can switch to vacuum boundary conditions $\Phi \to 0$ as $r' \to \infty$. To get the true gravitational field strength, we would simply need to add ${n_{ext}}z$. Switching boundary conditions in this way, we end up dealing with the potential due to the mass only \newline ($\equiv \Phi$ when the mass is present $- ~ \Phi$ when it is absent).

For a point mass, we expect that $\Phi \propto -\frac{1}{r'}$ where $r' \equiv \sqrt{x'^2 + y'^2 + z'^2}$. To find the normalisation, we note that
\begin{eqnarray}
	\int \nabla'^2 \Phi ~ d^3 \bm{r'} &=& \int \frac{4 \pi G \rho}{\mu_{ext}} ~ d^3 \bm{r'} \\
	&=& \int \frac{4 \pi G \rho}{\mu_{ext} \sqrt{1 + L_0}} ~ d^3 \bm{r} \\
	&=& \frac{4 \pi G M}{\mu_{ext} \sqrt{1 + L_0}} ~\text{ as } \int \rho ~ d^3 \bm{r} \equiv M
\end{eqnarray}

As a result, the solution must be
\begin{eqnarray}
	\Phi &=& - ~ \frac{GM}{\mu_{ext} ~r' \sqrt{1 + L_0} } \\
	&=& - ~ \frac{GM}{\mu_{ext} \sqrt{\left( 1 + L_0 \right)\left( x^2 + y^2 \right) + z^2}} \\
	&=& - ~ \frac{GM}{\mu_{ext} ~r \sqrt{1 + L_0 \sin^2 \theta}}
	\label{Primary_result_AQUAL}
\end{eqnarray}

In Equation \ref{Primary_result_AQUAL}, we defined $r$ analogously to $r'$ and used spherical polar co-ordinates such that $z \equiv r \cos \theta$ (see bottom panel of Figure \ref{Force_comparison}). This makes it easier to see an unusual aspect of the solution: gravity due to the mass is not always towards it. The components of the gravitational field strength $\bm{g}$ in the radial and tangential directions are
\begin{eqnarray}
	\bm{g}_{_r} &=& - ~ \frac{GM}{r^2 \mu_{ext} \sqrt{1 + L_0 \sin^2 \theta}} \\
	\bm{g}_{_\theta} &=& - ~ \frac{GML_0 \sin \theta \cos \theta}{r^2 \mu_{ext} {\left( 1 + L_0 \sin^2 \theta \right)}^{\frac{3}{2}}}
\end{eqnarray}

In the deep-MOND regime, $n \ll a_0$ and so $\mu \propto n$. As a result, $L_0 = 1$. Thus, at the same distance $r$ from the mass, $\bm{g}_{_r}$ is a factor of $\sqrt{2}$ smaller at positions orthogonal to $\bm{n}_{ext}$ compared with positions along it (Figure \ref{Force_comparison}). The difference vanishes if the external field is much stronger than $a_0$ because then $L_0 = 0$ and we recover Newtonian gravity.

\section{External Field Dominance In QUMOND}

Equation \ref{AQUAL_Governing_Equation} is difficult to solve numerically. This has led to the development of a new quasi-linear formulation of Modified Newtonian Dynamics \citep[QUMOND,][]{Milgrom_2010_QUMOND}. In this theory, one first has to obtain the Newtonian potential $\Phi_N$ associated with the matter distribution being solved for. An algebraic relation is applied to $\Phi_N$ to obtain the phantom dark matter density $\rho_{ph}$. The Newtonian potential of the actual plus phantom dark matter is the true potential $\Phi$. 
\begin{eqnarray}
	\nabla \cdot \left[ \nu \left( \left| \bm{n}_{_N} + \bm{n}_{_{N, ext}} \right| \right) ~ \left( \bm{n}_{_N} + \bm{n}_{_{N, ext}} \right)\right] ~=~ \nabla^2 \Phi
	\label{QUMOND_Governing_Equation}
\end{eqnarray}

Variables with a subscript $_N$ denote values in Newtonian gravity. It should be clear that we can solve the equation using vacuum boundary conditions and just add in a constant external field at the end. One has to be a little careful about the meaning of the external field in this case. We assume that the relation between the true and Newtonian external fields is the same as for a point mass.
\begin{eqnarray}
	\bm{n}_{ext} = \nu \left( \bm{n}_{_{N, ext}} \right) \bm{n}_{_{N, ext}}
\end{eqnarray}

The $\nu$ function must have the asymptotic limits $\nu \to 1$ for $n_{_N} \gg a_0$ while in the opposite limit, $\nu \to \sqrt{a_{_0}/n_{_N}}$. The $\nu$ function corresponding to the $\mu$ function in Equation \ref{Equation_simple_mu} is
\begin{eqnarray}
	\nu ~=~ \frac{1}{2} \left( 1 + \sqrt{1 + \frac{4 a_{_0}}{n_{_N}}} ~\right)
\end{eqnarray}

It can be verified straightforwardly that $\mu \left( g \right) \nu \left( g_{_N} \right) = 1$ if one obtains $g$ by implicitly solving $g \mu \left( g \right) = g_{_N}$ or from the explicit relation $g = g_{_N} \nu \left( g_{_N} \right)$.

Proceeding in a similar manner to our derivation for AQUAL, the QUMOND analogue of Equation \ref{AQUAL_linearised} is
\begin{eqnarray}
	\label{Equation_19}
	\nabla^2 \Phi ~&=&~ \nu_{ext} \left( \nabla \cdot \bm{n}_{_N} + K_0 \frac{\partial \bm{n}_{_{N, z}}}{\partial z} \right) ~~ \text{ where} \\
	K_0 &\equiv & \frac{\partial Ln ~ \nu}{\partial Ln ~ {n}_{_{N, ext}}} ~~ \text{ and} \\
	\bm{n}_{_N} ~&\equiv &~ \frac{GM\bm{r}}{r^3} ~~\left(\text{note this is } -\bm{g}_{_N} \right)
\end{eqnarray}

The structure of the equations is similar so far, except that $\mu$ usually appears in the denominator while $\nu$ appears in the numerator. However, we have only managed to determine $\nabla^2 \Phi$, not $\Phi$ itself. To find out what it is, we note that $\nabla \cdot \bm{n}_{_N} = 0$ everywhere except at the point mass. Thus, we expect that there will be a $\frac{1}{r}$ term in the final potential $\Phi$. We will determine the magnitude of this later.

First, we focus on the potential $\Phi_{smooth}$ due to the non-singular part of the effective matter density. $\Phi_{smooth}$ is sourced by the $\frac{\partial \bm{n}_{_{N, z}}}{\partial z}$ term in Equation \ref{Equation_19}. Using co-ordinates centred on the mass with $z$ along the external field direction (as before), we get that 
\begin{eqnarray}
	\frac{\partial \bm{n}_{_{N, z}}}{\partial z} ~&=&~ \frac{\partial}{\partial z} \left( \frac{GMz}{r^3}\right) \\
	~&=&~ \frac{GM}{r^3} \left(1 - 3 \cos^2 \theta \right)
\end{eqnarray}


We now use Equation 2.95 of \citet{Galactic_Dynamics} to obtain that
\begin{eqnarray}
	\Phi_{smooth} ~~=~~ \frac{GM\nu_{ext} K_0}{6r} \left( 3 \cos^2 \theta - 1\right)
	\label{Phi_smooth}
\end{eqnarray}

This can be verified by direct substitution into the Poisson Equation. A $\frac{1}{r}$ term in the potential does not contribute to the radial part of $\nabla^2 \Phi$. The angular part is the Legendre polynomial $P_L \left( \cos \theta \right) $, with $L = 2$. Its eigenvalue under the Laplacian operator is $-L \left( L + 1 \right) = -6$, hence the result.

We still need to determine the contribution to $\Phi$ from the regions very close to the mass. We call this $\Phi_{point}$. To find its magnitude, we will again calculate $\int \nabla^2 \Phi ~d^3 \bm{r}$. We must first obtain the contribution to this integral from $\Phi_{smooth}$. The rest is necessarily due to $\Phi_{point}$.
\begin{eqnarray}
	\int \nabla^2 \Phi_{smooth} ~ d^3 \bm{r} &=& \int \frac{\partial \Phi_{smooth}}{\partial r} ~ dS \\
	&\propto & \int_0^\pi \left(3 \cos^2 \theta - 1 \right) \sin \theta ~ d\theta \\
	&=& 0
\end{eqnarray}

We took advantage of the Divergence Theorem to convert a volume integral into a surface integral. We now integrate Equation \ref{Equation_19} over all space to get the normalisation of $\Phi_{point}$. Firstly, we note that
\begin{eqnarray}
	\int \nabla \cdot \bm{n}_{_N} ~ d^3 \bm{r} ~=~ 4 \pi G M
	\label{Equation_27}
\end{eqnarray}

This follows immediately from the normal Poisson Equation $\nabla^2 \Phi_{_N} = 4 \pi G \rho$. Due to rotational symmetry of $\bm{n}_{_N}$, this integral must receive identical contributions from $\frac{\partial^2 \Phi_N}{\partial x^2}$, $\frac{\partial^2 \Phi_N}{\partial y^2}$ and from $\frac{\partial^2 \Phi_N}{\partial z^2}$. Therefore, we must have that
\begin{eqnarray}
	\int \frac{\partial \bm{n}_{_{N, z}}}{\partial z} ~ d^3 \bm{r} ~=~ \frac{4 \pi G M}{3}
	\label{Equation_28}
\end{eqnarray}


Using Equations \ref{Equation_27} and \ref{Equation_28} in Equation \ref{Equation_19}, we get that
\begin{eqnarray}
	\Phi_{point} ~=~ - ~ \frac{GM \nu_{ext}}{r} \left( 1 + \frac{K_0}{3} \right)
\end{eqnarray}

Therefore, the total potential must be
\begin{eqnarray}
	\Phi ~=~ -\frac{GM \nu_{ext}}{r} \left( 1 + \frac{K_0}{2} \sin^2 \theta \right)
	\label{Primary_result_QUMOND}
\end{eqnarray}

The components of the gravitational field strength are
\begin{eqnarray}
	\bm{g}_r &=& - ~ \frac{GM \nu_{ext}}{r^2} \left( 1 + \frac{K_0}{2} \sin^2 \theta \right) \\
	\bm{g}_{_\theta} &=& \frac{GM \nu_{ext}}{r^2} \left( K_0 \sin \theta \cos \theta \right)
\end{eqnarray}

\begin{figure}
	\centering 
		\includegraphics [width = 8.5cm] {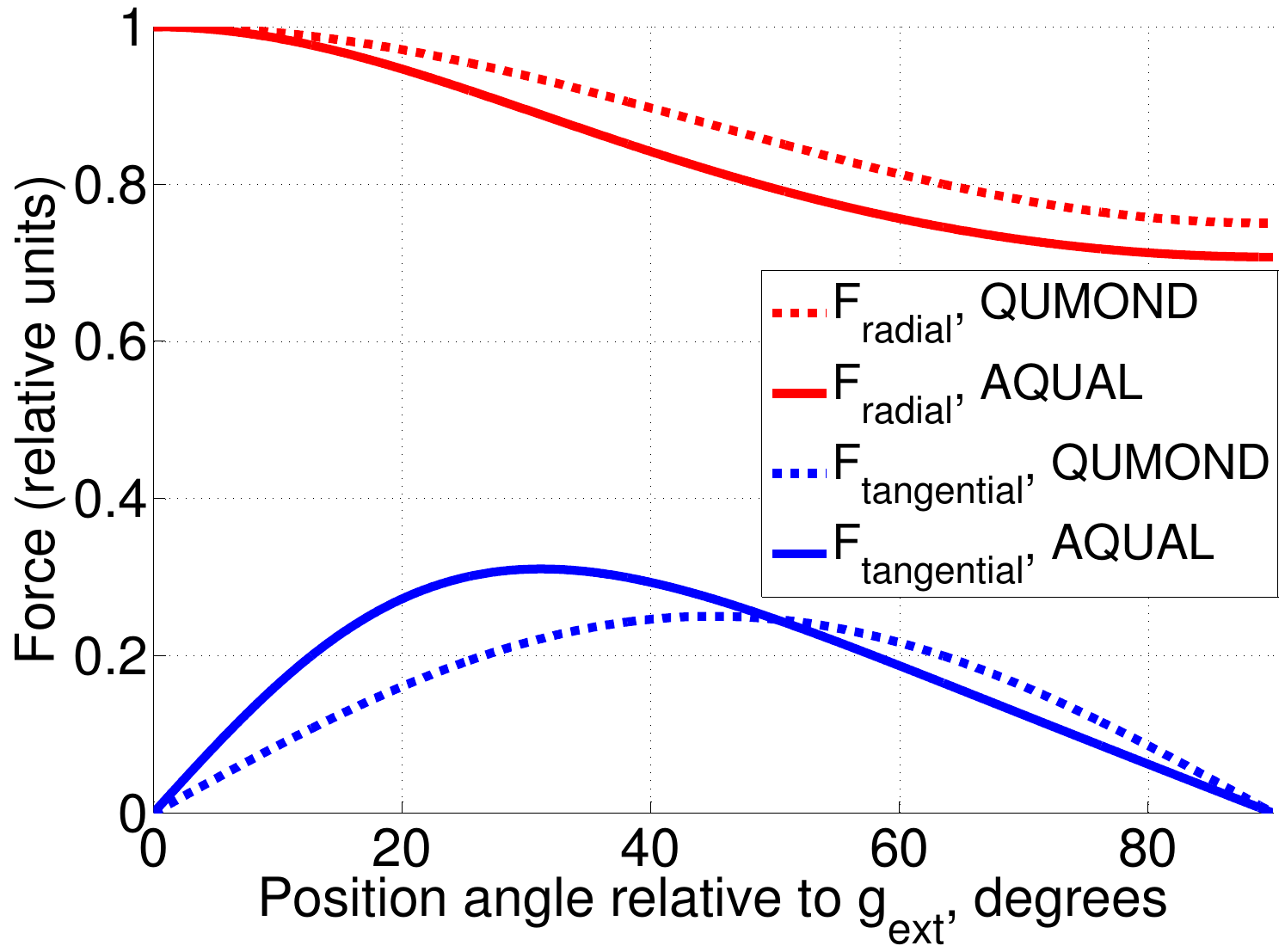}
		\includegraphics [width = 8.5cm] {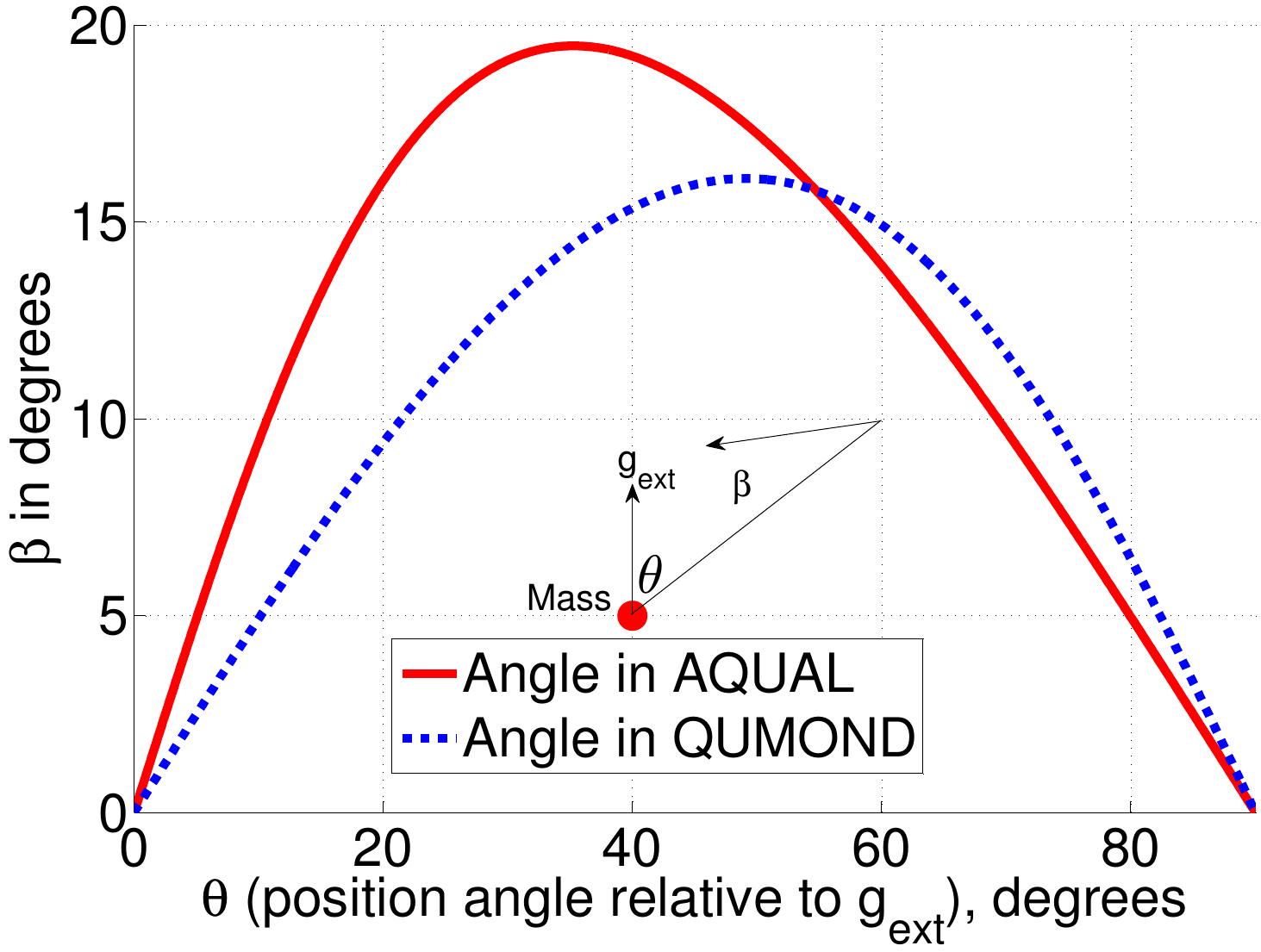}		
	\caption{\emph{Top:} The angular dependence of the radial and tangential components of the gravitational field are shown in the case of external field dominance in the deep-MOND limit (accelerations $\ll a_0$). Forces are in units of $\frac{GM \nu_{_{ext}}}{r^2}$ or $\frac{GM}{\mu_{_{ext}} r^2}$ (see text). \emph{Bottom:} The angle $\beta$ between the force and the radial direction. The inset figure shows the sense of $\beta$. The deepest parts of the potential are along the external field direction in both theories.}
	\label{Force_comparison}
\end{figure}

As in AQUAL, the force due to the mass is not always directly towards it (although it is never more than 20$^\circ$ off). In the deep-MOND limit, $n_{_{N, ext}} \ll a_0$ and so $\nu \propto 1/\sqrt{n_{_N}}$. As a result, $K_0 = -\frac{1}{2}$. This implies that, at the same distance $r$ from the mass, its gravitational field is $\frac{3}{4}$ as strong for points orthogonal to $\bm{n}_{ext}$ compared with points along it. The forces in AQUAL and QUMOND are compared in Figure \ref{Force_comparison}.

\section{Application to the Sagittarius Tidal Stream}
\label{Sgr_tidal_stream}

To see how motions might differ between AQUAL and QUMOND, we conducted a basic investigation of the Sgr tidal stream (parameters in Table \ref{Parameters}). Sgr was evolved in the potential of the MW, which we treated as an isolated point mass. Thus, the forces on Sgr are the same in QUMOND and AQUAL, leading to the same orbit.

We performed our simulation in 2D as Sgr would move within a plane. At $\approx 32$ points per orbit, we created 241 test particles at each of the Lagrange points L$_1$ and L$_2$. These particles had velocities relative to Sgr which covered the possible range of directions and had magnitudes 0 $-$ 3 times its velocity dispersion. L$_{1,2}$ are located on the MW$-$Sgr line where, in a reference frame co-rotating with the instantaneous angular velocity of Sgr, the combination of centrifugal and tidal forces from the MW first overcomes gravity from Sgr. More details can be found in \citet{Zhao_Tian_2006}.

We used an adaptive timestep procedure with forces as in the analytic solutions derived earlier (Equation \ref{Primary_result_AQUAL} or \ref{Primary_result_QUMOND}). To prevent the force from Sgr diverging close to its centre, we softened the force within a distance of $r_{core}$. We took this to be half the minimum distance from Sgr to L$_1$ over its entire orbit.

The tidal debris end up covering more than 360$^\circ$ around the MW. Thus, we used the concept of orbital phase angle. The idea should still work in 3D, at least for test particles that do not go too far outside the orbital plane of Sgr.

\begin{figure}
	\centering
		\includegraphics [width = 8.15cm] {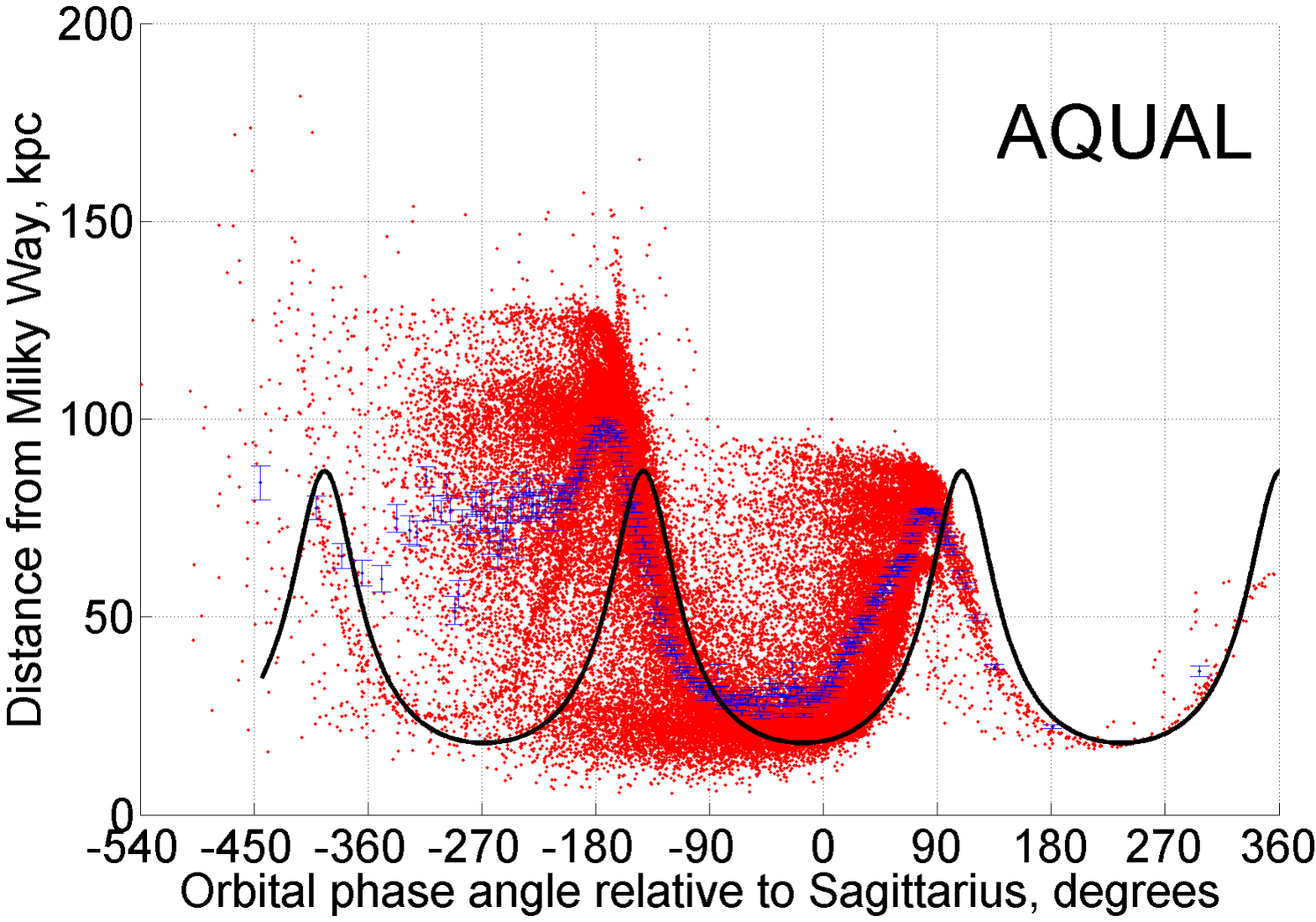}
		\includegraphics [width = 8.15cm] {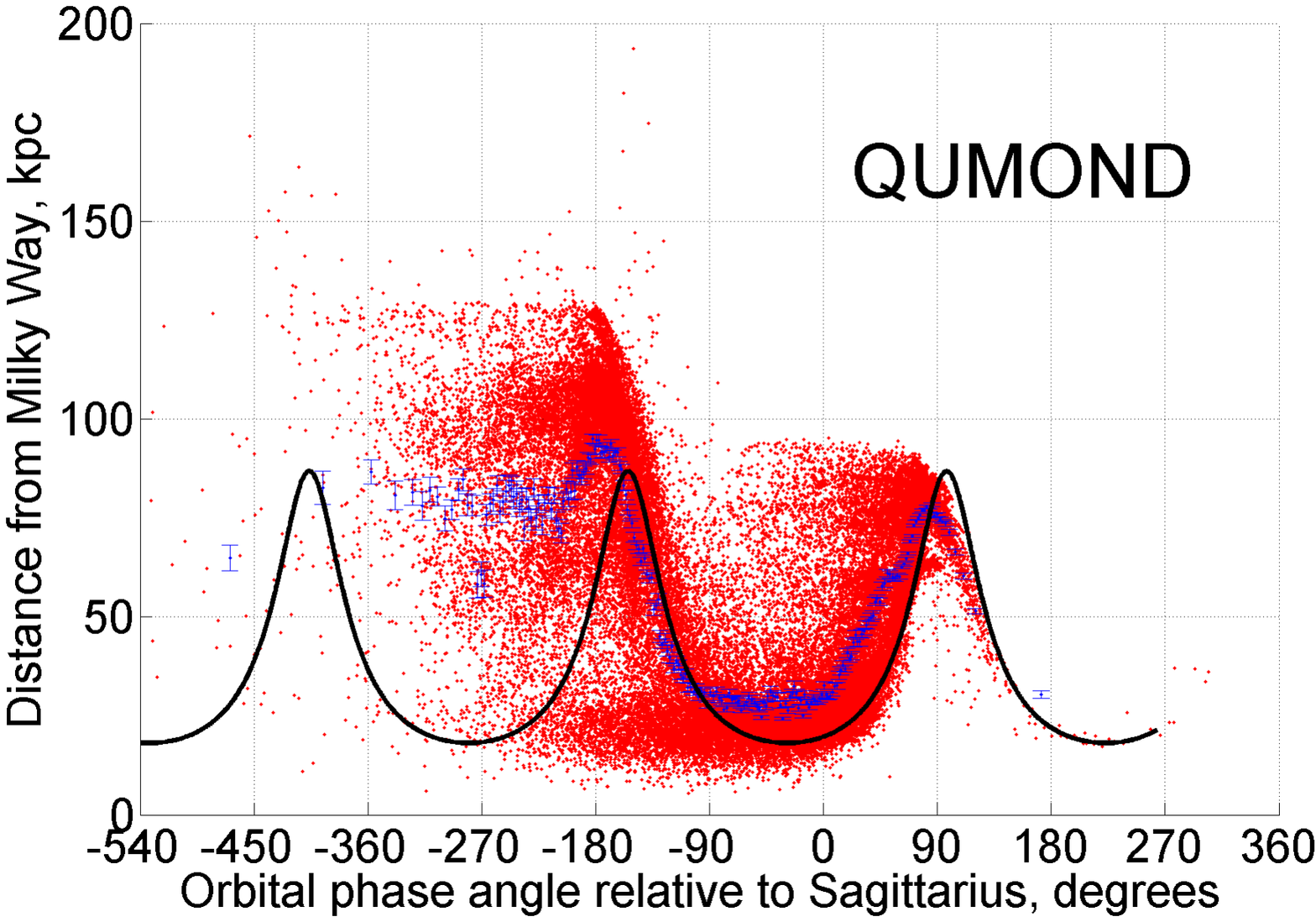}
		\caption{Orbital radius as a function of phase angles for test particles in our modified Lagrange Cloud stripping simulations of the Sgr tidal stream (red dots, error bars show binned results). Sgr is at 0$^\circ$ and its orbit is shown as a black line. Notice that AQUAL (\emph{top}) and QUMOND (\emph{bottom}) give very similar results.}
	\label{Tidal_stream_comparison}
\end{figure}

Our results are shown in Figure \ref{Tidal_stream_comparison}. It is apparent that there is almost no difference between AQUAL and QUMOND, despite very different-looking equations. Thus, it should be possible to determine parameters of the MW and Sgr in MOND using one of these theories and expect the results to be very similar in the other theory.

In this particular problem, $\nu_{ext} \approx 5$. Thus, the Sgr mass inferred from a Newtonian analysis of the data would be $\approx 5 \times$ its baryonic mass. Sgr would \emph{appear} to be dominated by dark matter, even if it had none.

\begin{table}
  \centering
  	\caption{Initial conditions for our Sgr tidal stream simulations.}
  	\begin{tabular}{cc}
   	\hline
	Parameter &	Value \\
	\hline
	Mass of Milky Way (MW) & $7 \times 10^{10} M_\odot$ \\
	Mass of Sagittarius (Sgr) & $10^8 M_\odot$ \\
	Internal velocity dispersion of Sgr & 9.85 km/s \\
	$\mu_\alpha \cos \delta$  (proper motion of Sgr in RA)& -2.56 mas/yr \\
	$\mu_\delta$ (proper motion of Sgr in declination) & -1.1884 mas/yr \\
	Present heliocentric distance to Sgr & 29.4 kpc \\
	Present heliocentric radial velocity of Sgr & -140 km/s \\
	\hline
	\end{tabular}
    \label{Parameters}	
\end{table}

\section{Conclusion}
We derived a new analytic result for the potential created by a point mass in QUMOND, in the case where an external field dominates the Newtonian gravitational field strength (Equation \ref{Primary_result_QUMOND}). We found that the forces are very similar to the original AQUAL formulation of MOND (see comparison in Figure \ref{Force_comparison}). To investigate further, we conducted a basic simulation of the Sagittarius tidal stream using the modified Lagrange Cloud Stripping procedure. The results are shown in Figure \ref{Tidal_stream_comparison}. Both formulations of MOND give almost identical results. This is due to the orbit of Sgr being the same in both cases and forces from Sgr being very similar. Thus, we expect that one can safely choose one formulation and expect the results to be very similar in the other (although we think the inferred $M_{Sgr}$ is slightly lower in AQUAL).

\section*{Acknowledgements}

IB is supported by a STFC studentship. He wishes to thank Rachel Cochrane for helpful comments.

\bibliographystyle{mnras}
\bibliography{DML_bbl}
\label{lastpage}
\end{document}